\documentclass{article}
\usepackage{spconf,amsmath,graphicx}

\ninept
\usepackage[english]{babel}

\usepackage{ifpdf}

\usepackage{cite} 
\usepackage{url}
\usepackage{hyperref}

\usepackage{color}
\usepackage{amssymb}
\usepackage{pgf}
\usepackage{pgf, tikz, pgfplots}
\usepackage{tikz-3dplot}
\usepackage{tikz-qtree}
\usetikzlibrary{shapes, arrows, automata, plotmarks}
\usetikzlibrary{calc,hobby,decorations,math}
\usepackage{tcolorbox}

\usepackage{amsfonts, amssymb, amsthm}
\usepackage{mathrsfs}

\usepackage{algorithm,algpseudocode}
	\algnewcommand{\LeftComment}[1]{\Statex \(\triangleright\) #1}
\newcommand{\End}[1]{\textrm{End}\left(#1\right)}

\usepackage{enumerate}
\usepackage{multirow}
\usepackage{rotating}
\usepackage{subcaption}
\usepackage{needspace}

\input{./latex_resources/mySymbol.sty}
\input{./latex_resources/pennColors.sty}

\newtheorem{definition}{\hspace{0pt}\bf Definition}

\title{Algebraic Convolutional Filters on Lie Group Algebras}
\name{Harshat Kumar, Alejandro Parada-Mayorga, and Alejandro Ribeiro}
\address{Department of Electrical and Systems Engineering, University of Pennsylvania, PA}
\begin{document}
%
\maketitle
\begin{abstract}
Group convolutional neural networks are a useful tool for utilizing symmetries known to be in a signal; however, they require that the signal is defined on the group itself. Existing approaches either work directly with group signals, or they impose a lifting step with heuristics to compute the convolution which can be computationally costly. Taking an algebraic signal processing perspective, we propose a novel convolutional filter from the Lie group algebra directly, thereby removing the need to lift altogether. Furthermore, we establish stability of the filter by drawing connections to multigraph signal processing. The proposed filter is evaluated on a classification problem on two datasets with $SO(3)$ group symmetries. 
\end{abstract}
%




\section{Introduction}
\vspace{-5pt}
The empirical success of convolutional neural networks (CNNs) and graph neural networks (GNNs) is generally attributed to their ability to capture inherent symmetries (translations and permutations respectively) in the signal \cite{bruna2013invariant, gama2020stability}. The desire to generalize this ability has led to the renewed investigation of group convolutions in group convolutional neural networks (G-CNNs), which are equivariant, i.e. the group action applied to a filtered signal equals the filter applied to the \emph{group actioned} signa) \cite{cohen2016group}. G-CNNs have been applied to a wide range of applications including images\cite{weiler2018learning}, point clouds\cite{esteves2018learning}, dynamical systems \cite{finzi2020generalizing}, and cancer detection in histopathology slides\cite{romero2020attentive}, where they are computed by transforming, or \emph{lifting}, the signal to be defined on the group. In this work, we generalize group convolutions by introducing algebraic convolutional filters for \emph{any} signal with a Lie group symmetry. 

Group convolutions have been implemented in a number of ways. Early approaches augment the CNN architecture by including copies of filters that have been acted on by group elements\cite{cohen2016group}. Building on this, information on the harmonics has been used to build filters which extend the prior approach to continuous groups \cite{worrall2017harmonic,esteves2018learning}. More recently, using the PointConv \cite{wu2019pointconv} method, Lie group convolutions have been considered for non-homogeneous spaces \cite{finzi2020generalizing}. While still requiring a group signal, they use the additional information of the orbits to codify a notion of locality. Here, a looser form of equivariance (in distribution only) can be guaranteed, and the signal still requires lifting if not defined on the group.

Through the lens of algebraic signal processing (ASP) \cite{puschel2008algebraic, parada2021algebraic}, we propose a novel group algebra convolution, of which group convolutions are a particular case. In particular, we consider Lie group symmetries in signals, which require the extension of the ASP model to Banach $*$-algebras. After the ideal filter is derived using the representation (or induced transformation) of the Lie group, we describe two approximation steps to realize a tractable filter. The first requires the sampling of the Lie group, while the second relies on the interpolation of the signal space. 

We have four main contributions. First, we define group convolutions when the domain of the signal is not a homogeneous space. For example, translation on pixels and rotations on angles are homogeneous, and our method allows for translation on angles and rotation on pixels. Second, we decouple the discretization of the Lie group convolution elucidating two separate sampling instances, namely the sampling on the input as well as the sampling on the group. Third, we provide an algebraically justified approach to approximate the group operator which can be done offline, enabling problems to scale. Finally, we show stability of our proposed signal by drawing connections to ASP on multigraphs\cite{zhang2018scalable,butler2022convolutional}. We evaluate the proposed filter numerically on the binary classification task of knots as well as on ModelNet10 \cite{wu20153d} (see Section \ref{sec_numericals}). 


\section{Algebraic Signal Processing for Lie Groups}
\vspace{-5pt}
In this section we introduce the framework that will allow us to do formal \textit{convolutional} processing of information on arbitrary spaces affected by the symmetries of a Lie group. 
The symmetries on a Lie group $G$ can be associated with transformations, or \textit{actions}, on the domain of spaces of functions. Formally, let $\ccalX\subset\mbR^{n}$ be a set and $f:\ccalX\to \mbC$ be a function. A Lie group $G$ acting on $\ccalX$ means that each element $g\in G$ has a corresponding transformation $T_g:\ccalX\to\ccalX$ called a group action, such that for any $g_1, g_2 \in G$, $T_{g_1 g_2} = T_{g_1}T_{g_2}$. Then, the group action on $\ccalX$ extends to $f$ by the transformation $\mbT_g: f \to \tilde{f}$, where $\tilde{f}(T_g(x)) = f(x)$ \cite{kondor2018generalization}. For example, consider the group of translations on the real line. In this case, $T_t(x) = x+t$ and $\tilde{f}(x) = f(x-t)$. If for any $x, y\in\ccalX$ there exists $g\in G$ such that $T_{g}y = x$, we call $\ccalX$ a homogeneous space. In this work, we will consider signals which have group symmetries, of which homogeneous spaces are a particular case.


The group convolutions we propose are an extension of the classical notion of Algebraic Signal Processing (ASP).
%
%
%
Namely, a signal model is defined by the triplet $(\ccalA, \ccalH, \rho)$, where $\ccalA$ is an associative algebra with unity, $\ccalH$ is a vector space, and $\rho:\ccalA \to \End{\ccalH}$ is a homomorphism between the algebra $\ccalA$ and the set of endomorphisms on $\ccalH$~\cite{puschel2006algebraic}. The vector space $\ccalH$ and the homomorphism $\rho$ together make a representation $(\ccalH, \rho)$ of the associative algebra $\ccalA$ by preserving multiplication and unit. Signals are modeled as the elements in $\ccalH$, while filters are elements of $\ccalA$. Using $\rho$ we instantiate the \textit{abstract} filters in $\ccalA$ into concrete operators in $\text{End}(\ccalH)$ that transform the signals in $\ccalH$. The ASP framework allows one to express a vast variety of convolutional models, including convolutional signal processing models on graphs, graphons, quivers, lattices, sets, and groups, among others~\cite{algSP2,algSP3,puschel_asplattice,puschel_aspsets,alejopm_gpooling_c,ParadaMayorga2020QuiverSP}.

We emphasize that ASP models encapsulate into algebraic principles the properties of a signal model, but by themselves they do not attach topological properties to any of the objects in the triplet $(\ccalA, \ccalH, \rho)$. Lie groups are not only algebraic objects, but also manifolds that link the differentiability and continuity of the manifold to the algebraic properties of the group. Therefore, in order to capture Lie group symmetries in general algebraic convolutional signal models, we need to extend ASP to include topological properties, which we describe in the following subsection.


\subsection{Extending ASP to Banach $*$-algebras}
\label{subsec_asp_BanachA}

To leverage ASP for the modeling of convolutional processing with/on Lie groups, we need to extend the classical notion of algebraic signal model (ASM) to include topological properties. We achieve this by exploiting the notions of Banach $\ast$-algebra, Hilbert space, and $\ast$-homomorphism~\cite{folland2016course,deitmar2014principles}. We recall that a Banach $\ast$-algebra, $\ccalA$, is an algebra which is also a Banach space and it is endowed with a closed operation $(\cdot)^\ast: \ccalA \rightarrow \ccalA$ called the adjoint. The adjoint must satisfy that $(ab)^\ast = b^\ast a^\ast$ for all $a,b\in\ccalA$, where $ab\in\ccalA$ indicates the product of $a$ and $b$. We also recall that $\ccalH$ is a Hilbert space if $\ccalH$ is endowed with an inner product, and it is complete as a metric space with respect to the metric induced by the inner product. If we denote by $\ccalB (\ccalH)$ the set of bounded operators acting on $\ccalH$, we say that $\rho: \ccalA \rightarrow \ccalB (\ccalH)$ is a $\ast$- homomorphism if $\rho$ is a homomorphism and $\rho\left( a^\ast \right) = \rho(a)^\ast$. With these concepts at hand, we now formalize the notion of ASM used in this paper.


\begin{definition}\label{def_ASM}
 
 An algebraic signal model (ASM) is a triplet $(\ccalA, \ccalH, \rho)$, where $\ccalA$ is a Banach $\ast$-algebra with unity, $\ccalH$ is a Hilbert space, and $\rho$ is a $\ast$-homomorphism between $\ccalA$ and $\ccalB (\ccalH)$.
 
\end{definition}


From Definition~\ref{def_ASM}, a rich variety of convolutional signal models can be derived. Fixing $\ccalH$ and choosing different algebras and/or homomorphisms, it is possible to leverage different types of symmetries of the data. Likewise, while fixing the algebra, it is possible to leverage specific symmetries into different types of signals and on different domains. We define the Lie group algebra formally. 

\begin{definition}[Lie group algebra]\label{def_LieGroupAlgebra} Given a group $G$ and Haar measure $\mu$,  the \textbf{Lie group algebra}, denoted $L^1(G)$, is defined by
%
%
%
 $L^1 (G)
       :=
       \left\lbrace 
               \boldsymbol{f}\in C_{c}(G) 
       \left\vert
                \int_G \vert \boldsymbol{f}\vert d\mu
                <\infty
       \right.            
       \right\rbrace
       .$
\end{definition}

For our discussion and for the rest of the paper we choose $\ccalA = L^1 (G)$ for a given Lie group $G$. 
%
%
%
The vector space $L^{1}(G)$ becomes an algebra by choosing a product $\ast$ between elements of $L^{1}(G)$ given by the well known group convolution \cite{folland2016course,deitmar2014principles}.
%
%
Now, we define on $L^{1}(G)$ an involution operation given by
$\boldsymbol{f}^{\ast}(x) = \Delta (x^{-1})\overline{\boldsymbol{f}(x^{-1})}$, where $\Delta$ is the modular function of $G$. In particular, if $\mu$ is a left Haar measure on the group $G$ and $x\in G$, we can define another measure $\mu_{x}(E) = \mu (Ex)$. As shown in~\cite{folland2016course} there exists a unique function $\Delta (x)$ -- independent of $\mu$ -- such that $\mu_{x} = \Delta(x)\mu$. The term $\Delta: G \rightarrow (0,\infty)$ is known as the modular function of $G$. Putting all these attributes together, $L^{1}(G)$ is a Banach $\ast$-algebra.

%
%
%
%
%
%

\begin{figure}
        \centering
        \input{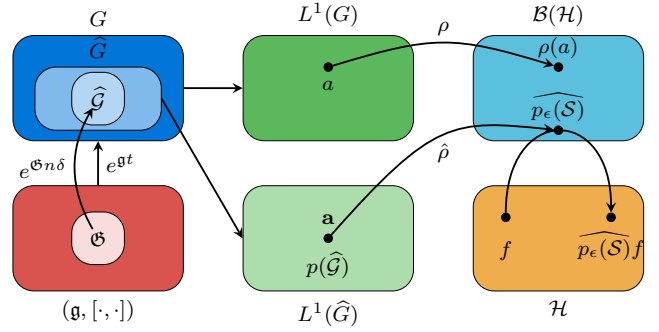} 
        \caption{Lie Algebraic Signal Model. The algebraic filters $a$ lie in the Lie group algebra $L^1(G)$, which are realized through the homomorphism $\rho$ which acts on the signals in the Hilbert space $\ccalH$ by \eqref{equ_rho_ideal}. The Lie algebra with bracket $(\mathfrak{g}, [\cdot, \cdot])$ is used with the exponential map to obtain a sampling of the Lie Group in order to approximate the ideal filter $a$ with a tractable counterpart $\mathbf{a}$.  }
        \label{fig:ASP_model}
\end{figure}

Given the induced group action $\mbT_g$, we let $\rho: L^1(G)\to \ccalB(\ccalH)$ be the \emph{algebra homomorphism} by letting 
\begin{equation} \label{equ_rho_ideal}
    \rho(a)f = \int_G a(g)\mbT_g f d\mu(g).
\end{equation}
The convolution expression in \eqref{equ_rho_ideal} has two main advantages. First, the operation is a formal convolution between the algebraic filter $a$ and the signal $f\in\ccalH$. This convolutional operation is not conditioned to have the elements in $\ccalH$ as functions whose domain is a homogeneous space. Second, the induced group action $\mbT_g$ is completely determined by the geometry of the space $\ccalX$ and can be computed offline. The model is shown in Figure \ref{fig:ASP_model}. Also shown in Figure \ref{fig:ASP_model} are the steps to make the integral in \eqref{equ_rho_ideal} tractable, which we formally describe in the following section.

\section{Algebraic Relaxations} 
\label{sec:generators}
\vspace{-5pt}

There are two main steps to approximate the integral in \eqref{equ_rho_ideal}. The first is the sampling on the group, and the second is the sampling on the Hilbert space (which is generally where the signal lies in practical applications). 

\subsection{Sampling of the Lie Group} \label{sec_sampling_lie_group}
We first consider finding a representative sampling of $\hat G$ of $G$ so that we can construct filters on the algebra $\ccalM = \{\bba:\hat G\to \mbC\} = \{\sum_{g\in \hat G}\bba(g)g\}$. The homomorphism $\hat \rho$ is consequently given by 
\begin{equation} \label{equ_rho_discrete}
    \hat \rho(\bba)f = \sum_{g\in \hat G} \bba(g) \mbT_g f.
\end{equation}
In order to obtain $\hat G$, we turn to the corresponding Lie algebra, and we use a discrete exponential map to generate this discrete version of the group. We formalize this in the following definition.



\begin{definition}\label{def_Ghat} 

Let $G$ be a Lie group and let $\mathfrak{g}$ be its associated Lie algebra. Let $\mathfrak{G}\subset\mathfrak{g}$ be a basis of $\mathfrak{g}$ as a vector space, and let $\widehat{G}_\delta$ be the subset of $G$ given by
$\widehat{G}_{\delta}
         =
         \left\lbrace
              \left.
              e^{\delta n \mathfrak{X}} 
              \right\vert \mathfrak{X}\in\mathfrak{G}, n\in\mbZ, \delta\in\mbR
         \right\rbrace$
%
%
%
%
%
%
%
where $\delta$ is fixed. We say that $\widehat{G}_{\delta}$ is a discrete approximation of $G$ with resolution $\delta$.
\end{definition}


Now, we use the elements in $\widehat{G}_{\delta}$ to generate an approximated version of $L^{1}(G)$, that we denote by $L^{1}(\widehat{G}_{\delta})$. 
It is important to remark that the approximation proposed in Definition~\ref{def_Ghat} is justified by the fact that the exponential map is \emph{locally} one-to-one and onto (see~\cite[p.~67]{hall_liealg}). 

\subsection{Sampling the Hilbert space}
The second approximation takes the form of interpolation error from the sampled signal. Consider first a continuous signal $(\ccalX, f(\ccalX))$ such that the group transformation $\mbT_g f$ is well defined. In most applications of interest, the continuous signals are sampled according to some distribution. Define $\ccalX$ by the $N$ discrete points $X = \left\{x_i\in  \ccalX\right\}_{i = 0}^{N -1}$. We can similarly define a function $f_X:X \to \mbC$ by evaluating $f$ at $x_i \in X$ for all $i = 0, \dots, N-1$. When we apply the group transformation $T_g(x_i)$, it is not necessary that $T_g(x_i)\in X$. It follows that the induced group transformation on $X$ is not defined. However, it \emph{does} hold that $T_g(x_i)\in \ccalX$. Using this, we propose the use of interpolation to estimate the induced group transformation. In this section, we describe the algebraic justification for interpolation. We use $\bbf$ to denote the sampled signal.



Starting with the sampling $X = \left\{x_i \in \ccalX\right\}_{i = 0}^{N-1}$, we define the Voronoi cells around each point $x_i$ by using the inner product norm of the Hilbert space $d(a, b) := \sqrt{\langle a-b, a - b\rangle}$ by
%
 $   \ccalC_i := \left\{x\in \ccalX \vert d(x, x_i) \leq d(x, x_j) ~\textrm{for all}~ i \neq j\right\}.$
%
Given the Voronoi cells $\ccalC_i$, for $i = 0, \dots N-1$, we define the subspace of piece-wise constant functions on the Voronoi cells
%
$    \ccalF:= \{f\in \ccalH | f(x)= c_i \in \mbC $$~\textrm{for all}~ x \in \ccalC_i\}.$

Consider the isomorphism between $\ccalF$ and $f_X$, $\xi:\ccalF \to f_X$, where the value of $f_X(x_i)$ is assigned by $c_i$. Further, recall that the induced group transformation is the image of an element in the Lie group algebra $\rho(a) = \mbT \in \ccalB\left(\ccalH\right)$. As such, we would ideally like the induced transformation on $f_X$ by 
$
    \rho_\textrm{ideal}(a) \equiv \xi \circ \rho(a) \circ \xi^{-1}.
$
Of course, as discussed earlier, it is not necessary that $\rho(a)\circ \xi^{-1} \in \ccalB(\ccalF)$. As such, we require a projection operator $\iota$ which projects any element of $\ccalH$ to the piece-wise constant subspace $\ccalF$. We define this mapping as $\iota: \ccalH \to \ccalF$. The resulting induced transformation therefore becomes  
$    \rho_\textrm{interp}(a) \equiv \xi \circ \iota \circ  \rho(a) \circ \xi^{-1}.
$
Due to the interpolation, there is a loss of information in $\rho_\textrm{interp}$. We note, however, that this error is minimized by the choice of interpolation schemes, which are well studied both for scattered points as well as structured grids \cite{mccarthy1992geometric,ameur2002interpolation}. In this work, we consider natural neighbor interpolation, which is known to be equivalent to barycentric interpolation \cite{bobach2009natural} shown in Algorithm \ref{alg:BarycentricInterpolator}. The proposed filter in \eqref{equ_rho_discrete} therefore becomes 
\begin{equation} \label{equ_proposed_filter}
    \Hat{\Hat{\rho}}(\bba)\bbf = \sum_{g\in \hat G} \bba(g) \hat \mbT_g \bbf = \sum_{k = 0}^{K-1}\bba_k \hat \mbT_{g(k)}\bbf,
\end{equation}
where we number the coefficients and group actions by $k=\{0, \dots, K-1\}$ given an ordering on the monomials. For example, given two generators $\{g_1, g_2\}$, define $\Pi_{1,2}(g_1, g_2) =\{g_1g_2^2, g_2g_1g_2, g_2^2g_1\}$, where the subscript denotes the number of times the $g_i$ generator appears. We denote the $k^\textrm{th}$ element of the set $\{\Pi_{j_1,j_2}(g_1, g_2)\}_{j_1, j_2 = 0}$ by $g(k)$. This extends to any finite number of generators. Note that $\Hat{\Hat{\rho}}$ depicts the estimation by relaxing the continuous group through the exponential map (section \ref{sec_sampling_lie_group}), and $\hat \mbT$ shows the estimation of the interpolation scheme.

\subsection{Stability of the relaxed filter}

Closer inspection of the realized filter \eqref{equ_proposed_filter} elucidates connections between the discretized Lie group signal and multigraphs \cite{knyazev2018spectral}. In particular, $\bbf$ can be viewed as a graph signal where each element $x_i$ is a node of the graph. Moreover, the sparse transformation $\hat \mbT_g$ takes the form of a graph shift operator. The image of the exponential map of each element in the Lie algebra basis $\mathfrak{X}\in \mathfrak{G}$ represents a separate graph shift operator whose combinations are not necessarily commutative. Let $g_i := e^{\delta n \mathfrak{X}}\vert_{n = 1}$. Then, we denote the set $\{g_1, \dots, g_n\}$ by $\widehat{\ccalG}$. Using this, the ASP model on multigraphs has the same convolutional filter form as \eqref{equ_proposed_filter} with two main differences. First, the algebra $\ccalA$ for multigraphs is the polynomial algebra with $n$ generators $t_1, \dots, t_n$. Second, the homomorphism is given by the shift operators $\rho(t_i) = \mbT_{g_i}$ constructed by Algorithm \ref{alg:Transformation}. In Figure \ref{fig:ASP_model}, we see the equivalence of $\bba$ and $p(\widehat{G})$, the polynomials on $\widehat{G}$.

Given that the realized filter is a non commutative ASP model on multigraphs, it follows that the stability results of non commutative algebras hold as well. In particular, by Corollary 1 of \cite{parada2021convolutional}, the stability (see \cite[Definition 10]{parada2021convolutional}) takes the from of 
\begin{equation} \label{equ_stability}
    \|p(\hat \mbT)\bbf - p(\ccalT(\hat \mbT))\bbf\| \leq C \left(\|\ccalT\|_\textrm{Lip} + O(\|\ccalT\|)\right)\|\bbf\|,
\end{equation}
where $\|\cdot \|_\textrm{Lip}$ is the Lipshitz norm and $\ccalT$ is a perturbation model.  The inequality in \eqref{equ_stability} states that a filter is stable when the deformation in the operator is proportional to the size of the deformation. 

\begin{algorithm}[t]
\caption{Offline transformation operator generation}
\begin{algorithmic}[1]
\Require Group action $g$, Interpolation $\iota$, sampled points $X$
\State Initialize sparse matrix $\hat \mbT \in \mbK^{N \times N}$
\State Apply the transformation $T_g = \rho_\textrm{ideal}$ to all $x_i$, $i = 0, \dots, N-1$
\For{$i = 0, \dots, n -1$}
\State $N, W \leftarrow$ \textbf{InterpScheme}$(x_i, T_g \circ X)$
\For{Neighbor $n_j\in N(x_i)$ }
\State Assign $\hat \mbT_{i,n_j} = W_j$
\EndFor
\EndFor
\State\Return $\hat \mbT$
\end{algorithmic}
\label{alg:Transformation}
\end{algorithm}
\begin{algorithm}[t]
\caption{InterpScheme: Barycentric Interpolation}
\begin{algorithmic}[1]
\Require Query $x_i$, Transformed points $TX$, $\#$ of neighbors $k$
\State Compute the distance of $x_i$ to each point in $TX$
\State Find indices of the $k$ smallest distances $n_0, \dots, n_{k-1}$ 
\State Solve the system of equations for $\pmb{\lambda}$
$$\begin{bmatrix}
1 & 1 & 1 & \dots & 1\\
x_{n_0}^0 & x_{n_1}^0 & x_{n_2}^0 & \dots & x_{n_{k-1}}^0 \\
\vdots & \vdots & \vdots & \ddots & \vdots\\
x_{n_0}^p & x_{n_1}^p & x_{n_2}^p & \dots & x_{n_{k-1}}^p \\
\end{bmatrix}
\begin{bmatrix}
\lambda_0 \\ \lambda_1 \\ \vdots \\ \lambda_{k-1} 
\end{bmatrix}
=
\begin{bmatrix}
x_i^0 \\ x_i^1 \\ \vdots \\ x_i^p
\end{bmatrix}
$$
\State \Return neighbors $n_{0, \dots, k-1}$ and weights $\pmb{\lambda}$
\end{algorithmic}
\label{alg:BarycentricInterpolator}
\end{algorithm}



\def\pathresults{./figures}

\begin{figure}[!ht]
\centering
	\begin{subfigure}{.5\textwidth}
	\centering



\definecolor{my_cp5_col6}{RGB}{2, 117, 216}
\definecolor{my_cp5_col5}{RGB}{92, 184, 92}
\definecolor{my_cp5_col4}{RGB}{91, 192, 222}
\definecolor{my_cp5_col3}{RGB}{240, 173, 78}
\definecolor{my_cp5_col2}{RGB}{217, 83, 79}
\definecolor{my_cp5_col1}{RGB}{52, 167, 255}

\usetikzlibrary{positioning,decorations.pathreplacing,shapes}


\def \scale {1}
\def \unit { \scale cm}

\def \vertsep {0.5*\scale}

\def \horzsep {1*\scale}


\tikzstyle{set} = [rectangle,color=black,
                    rounded corners = 0*\unit,
                    fill=black,
                    inner sep=0pt,
                    draw,
                    anchor = center,
                    line width=0.1mm]
                    
 
 
 \tikzstyle{myboxlabel} = [set,
fill=my_cp5_col3,
minimum width  = 0.485*\unit,
minimum height = 0.3*\unit]

\tikzstyle{dot} = [ circle,
                    minimum width  = 0.05*\unit,
                    fill=black,
                    color=black,
                    inner sep=0pt,
                    draw,
                    anchor = center ]


{\fontsize{8}{8}\selectfont

\begin{tikzpicture}[scale=\scale,rounded corners,ultra thick]


   
   \path (0,0) node [myboxlabel,fill=my_cp5_col6] (L1) {};
   \path (L1.south) ++ (0, 0) node [below, color=black] {GrpA-1};

   
    \path (L1.east)++(\horzsep,0) node [myboxlabel,fill=my_cp5_col5] (L2) {};
    \path (L2.south) ++ (0, 0) node [below, color=black] {GrpA-2};

     
     \path (L2.east)++(\horzsep,0) node [myboxlabel,fill=my_cp5_col4] (L3) {};
     \path (L3.south) ++ (0, 0) node [below, color=black] {FCNN-1};

         
     \path (L3.east)++(\horzsep,0) node [myboxlabel,fill=my_cp5_col3] (L4) {};
     \path (L4.south) ++ (0, 0) node [below, color=black] {FCNN-2};

      
      \path (L4.east)++(\horzsep,0) node [myboxlabel,fill=my_cp5_col2] (L5) {};
      \path (L5.south) ++ (0, 0) node [below, color=black] {LieConv-1};

      
      \path (L5.east)++(\horzsep,0) node [myboxlabel,fill=my_cp5_col1] (L6) {};
      \path (L6.south) ++ (0, 0) node [below, color=black] {LieConv-2};

%
%

\end{tikzpicture}

}
\end{subfigure}
	\centering
	\begin{subfigure}{.23\textwidth}
		\centering
		\resizebox{\linewidth}{!}{
\begin{tikzpicture}

\definecolor{my_cp5_col6}{RGB}{2, 117, 216}
\definecolor{my_cp5_col5}{RGB}{92, 184, 92}
\definecolor{my_cp5_col4}{RGB}{91, 192, 222}
\definecolor{my_cp5_col3}{RGB}{240, 173, 78}
\definecolor{my_cp5_col2}{RGB}{217, 83, 79}
\definecolor{my_cp5_col1}{RGB}{52, 167, 255}
\definecolor{burlywood231187113}{RGB}{231,187,113}
\definecolor{darkgray158150204}{RGB}{158,150,204}
\definecolor{darkslategray66}{RGB}{66,66,66}
\definecolor{dimgray85}{RGB}{85,85,85}
\definecolor{gainsboro229}{RGB}{229,229,229}
\definecolor{gray119}{RGB}{119,119,119}
\definecolor{indianred2049072}{RGB}{204,90,72}
\definecolor{steelblue69133171}{RGB}{69,133,171}
\definecolor{yellowgreen13817181}{RGB}{138,171,81}

\begin{axis}[
axis background/.style={fill=gainsboro229},
axis line style={white},
tick align=outside,
x grid style={white},
xmajorticks=false,
xmin=-0.5, xmax=5.5,
xtick style={color=dimgray85},
xtick={0,1,2,3,4,5},
xticklabels={GrpA-1,GrpA-2,FCNN-1,FCNN-2,LieConv-1,LieConv-2},
y grid style={white},
 ylabel=\textcolor{dimgray85}{\LARGE Sphere},
ymajorgrids,
ymin=0.4, ymax=0.9,
ytick pos=left,
ytick style={color=dimgray85}
]
\draw[draw=none,fill=my_cp5_col6,very thin] (axis cs:-0.4,0) rectangle (axis cs:0.4,0.695744680851064);
\draw[draw=none,fill=my_cp5_col5,very thin] (axis cs:0.6,0) rectangle (axis cs:1.4,0.715602836879433);
\draw[draw=none,fill=my_cp5_col4,very thin] (axis cs:1.6,0) rectangle (axis cs:2.4,0.636170212765958);
\draw[draw=none,fill=my_cp5_col3,very thin] (axis cs:2.6,0) rectangle (axis cs:3.4,0.631914893617021);
\draw[draw=none,fill=my_cp5_col2,very thin] (axis cs:3.6,0) rectangle (axis cs:4.4,0.708510638297872);
\draw[draw=none,fill=my_cp5_col1,very thin] (axis cs:4.6,0) rectangle (axis cs:5.4,0.702836879432624);
\addplot [line width=1.08pt, darkslategray66]
table {%
0 0.678723404255319
0 0.712056737588652
};
\addplot [line width=1.08pt, darkslategray66]
table {%
1 0.693617021276596
1 0.739716312056738
};
\addplot [line width=1.08pt, darkslategray66]
table {%
2 0.582978723404255
2 0.677304964539007
};
\addplot [line width=1.08pt, darkslategray66]
table {%
3 0.565957446808511
3 0.686524822695036
};
\addplot [line width=1.08pt, darkslategray66]
table {%
4 0.686524822695035
4 0.729078014184397
};
\addplot [line width=1.08pt, darkslategray66]
table {%
5 0.690780141843972
5 0.717021276595745
};
\end{axis}

\end{tikzpicture}}


  \resizebox{\linewidth}{!}{
\begin{tikzpicture}
\definecolor{my_cp5_col6}{RGB}{2, 117, 216}
\definecolor{my_cp5_col5}{RGB}{92, 184, 92}
\definecolor{my_cp5_col4}{RGB}{91, 192, 222}
\definecolor{my_cp5_col3}{RGB}{240, 173, 78}
\definecolor{my_cp5_col2}{RGB}{217, 83, 79}
\definecolor{my_cp5_col1}{RGB}{52, 167, 255}
\definecolor{burlywood231187113}{RGB}{231,187,113}
\definecolor{darkgray158150204}{RGB}{158,150,204}
\definecolor{darkslategray66}{RGB}{66,66,66}
\definecolor{dimgray85}{RGB}{85,85,85}
\definecolor{gainsboro229}{RGB}{229,229,229}
\definecolor{gray119}{RGB}{119,119,119}
\definecolor{indianred2049072}{RGB}{204,90,72}
\definecolor{steelblue69133171}{RGB}{69,133,171}
\definecolor{yellowgreen13817181}{RGB}{138,171,81}

\begin{axis}[
axis background/.style={fill=gainsboro229},
axis line style={white},
tick align=outside,
x grid style={white},
xmajorticks=false,
xmin=-0.5, xmax=5.5,
xtick style={color=dimgray85},
xtick={0,1,2,3,4,5},
xticklabels={GrpA-1,GrpA-2,FCNN-1,FCNN-2,LieConv-1,LieConv-2},
y grid style={white},
 ylabel=\textcolor{dimgray85}{\LARGE Uniform},
ymajorgrids,
ymin=0.4, ymax=0.9,
ytick pos=left,
ytick style={color=dimgray85}
]
\draw[draw=none,fill=my_cp5_col6,very thin] (axis cs:-0.4,0) rectangle (axis cs:0.4,0.653900709219858);
\draw[draw=none,fill=my_cp5_col5,very thin] (axis cs:0.6,0) rectangle (axis cs:1.4,0.65531914893617);
\draw[draw=none,fill=my_cp5_col4,very thin] (axis cs:1.6,0) rectangle (axis cs:2.4,0.599290780141844);
\draw[draw=none,fill=my_cp5_col3,very thin] (axis cs:2.6,0) rectangle (axis cs:3.4,0.648936170212766);
\draw[draw=none,fill=my_cp5_col2,very thin] (axis cs:3.6,0) rectangle (axis cs:4.4,0.669503546099291);
\draw[draw=none,fill=my_cp5_col1,very thin] (axis cs:4.6,0) rectangle (axis cs:5.4,0.631205673758865);
\addplot [line width=1.08pt, darkslategray66]
table {%
0 0.640425531914894
0 0.668794326241135
};
\addplot [line width=1.08pt, darkslategray66]
table {%
1 0.625514184397163
1 0.685833333333333
};
\addplot [line width=1.08pt, darkslategray66]
table {%
2 0.570212765957447
2 0.621985815602837
};
\addplot [line width=1.08pt, darkslategray66]
table {%
3 0.604255319148936
3 0.685124113475177
};
\addplot [line width=1.08pt, darkslategray66]
table {%
4 0.642553191489362
4 0.697163120567376
};
\addplot [line width=1.08pt, darkslategray66]
table {%
5 0.599290780141844
5 0.665975177304964
};
\end{axis}

\end{tikzpicture}}

\resizebox{\linewidth}{!}{
\begin{tikzpicture}

\definecolor{my_cp5_col6}{RGB}{2, 117, 216}
\definecolor{my_cp5_col5}{RGB}{92, 184, 92}
\definecolor{my_cp5_col4}{RGB}{91, 192, 222}
\definecolor{my_cp5_col3}{RGB}{240, 173, 78}
\definecolor{my_cp5_col2}{RGB}{217, 83, 79}
\definecolor{my_cp5_col1}{RGB}{52, 167, 255}
\definecolor{burlywood231187113}{RGB}{231,187,113}
\definecolor{darkgray158150204}{RGB}{158,150,204}
\definecolor{darkslategray66}{RGB}{66,66,66}
\definecolor{dimgray85}{RGB}{85,85,85}
\definecolor{gainsboro229}{RGB}{229,229,229}
\definecolor{gray119}{RGB}{119,119,119}
\definecolor{indianred2049072}{RGB}{204,90,72}
\definecolor{steelblue69133171}{RGB}{69,133,171}
\definecolor{yellowgreen13817181}{RGB}{138,171,81}

\begin{axis}[
axis background/.style={fill=gainsboro229},
axis line style={white},
tick align=outside,
x grid style={white},
xmajorticks=false,
xmin=-0.5, xmax=5.5,
xmajorticks=false,
xtick style={color=dimgray85},
xtick={0,1,2,3,4,5},
xticklabels={GrpA-1,GrpA-2,FCNN-1,FCNN-2,LieConv-1,LieConv-2},
y grid style={white},
ylabel=\textcolor{dimgray85}{\LARGE Grid},
ymajorgrids,
ymin=0.4, ymax=0.9,
ytick pos=left,
ytick style={color=dimgray85}
]
\draw[draw=none,fill=my_cp5_col6,very thin] (axis cs:-0.4,0) rectangle (axis cs:0.4,0.665957446808511);
\draw[draw=none,fill=my_cp5_col5,very thin] (axis cs:0.6,0) rectangle (axis cs:1.4,0.683687943262411);
\draw[draw=none,fill=my_cp5_col4,very thin] (axis cs:1.6,0) rectangle (axis cs:2.4,0.56241134751773);
\draw[draw=none,fill=my_cp5_col3,very thin] (axis cs:2.6,0) rectangle (axis cs:3.4,0.638297872340426);
\draw[draw=none,fill=my_cp5_col2,very thin] (axis cs:3.6,0) rectangle (axis cs:4.4,0.640425531914894);
\draw[draw=none,fill=my_cp5_col1,very thin] (axis cs:4.6,0) rectangle (axis cs:5.4,0.661702127659575);
\addplot [line width=1.08pt, darkslategray66]
table {%
0 0.624822695035461
0 0.71063829787234
};
\addplot [line width=1.08pt, darkslategray66]
table {%
1 0.646099290780142
1 0.722695035460993
};
\addplot [line width=1.08pt, darkslategray66]
table {%
2 0.537588652482269
2 0.582978723404255
};
\addplot [line width=1.08pt, darkslategray66]
table {%
3 0.590780141843972
3 0.670921985815603
};
\addplot [line width=1.08pt, darkslategray66]
table {%
4 0.613475177304964
4 0.672340425531915
};
\addplot [line width=1.08pt, darkslategray66]
table {%
5 0.631914893617021
5 0.690780141843972
};
\end{axis}

\end{tikzpicture}}
		
	\end{subfigure}
	\begin{subfigure}{.23\textwidth}
		\centering
%
\resizebox{\linewidth}{!}{
\begin{tikzpicture}

\definecolor{my_cp5_col6}{RGB}{2, 117, 216}
\definecolor{my_cp5_col5}{RGB}{92, 184, 92}
\definecolor{my_cp5_col4}{RGB}{91, 192, 222}
\definecolor{my_cp5_col3}{RGB}{240, 173, 78}
\definecolor{my_cp5_col2}{RGB}{217, 83, 79}
\definecolor{my_cp5_col1}{RGB}{52, 167, 255}
\definecolor{darkgray158150204}{RGB}{158,150,204}
\definecolor{darkslategray66}{RGB}{66,66,66}
\definecolor{dimgray85}{RGB}{85,85,85}
\definecolor{gainsboro229}{RGB}{229,229,229}
\definecolor{gray119}{RGB}{119,119,119}
\definecolor{indianred2049072}{RGB}{204,90,72}
\definecolor{steelblue69133171}{RGB}{69,133,171}

\begin{axis}[
axis background/.style={fill=gainsboro229},
axis line style={white},
tick align=outside,
tick pos=left,
x grid style={white},
xmin=-0.5, xmax=3.5,
xmajorticks=false,
xtick style={color=dimgray85},
xtick={0,1,2,3},
xticklabels={\Large GrpA-1,\Large GrpA-2,\Large FCNN-1,\Large FCNN-2},
y grid style={white},
 ylabel=\textcolor{dimgray85}{ \LARGE \textcolor{white}{Sphere}},
ymajorgrids,
ymin=0.4, ymax=0.9,
ytick style={color=dimgray85}
]
\draw[draw=none,fill=my_cp5_col6,very thin] (axis cs:-0.4,0) rectangle (axis cs:0.4,0.763931104356636);
\draw[draw=none,fill=my_cp5_col5,very thin] (axis cs:0.6,0) rectangle (axis cs:1.4,0.765957446808511);
\draw[draw=none,fill=my_cp5_col4,very thin] (axis cs:1.6,0) rectangle (axis cs:2.4,0.646572104018913);
\draw[draw=none,fill=my_cp5_col3,very thin] (axis cs:2.6,0) rectangle (axis cs:3.4,0.67612293144208);
\addplot [line width=1.08pt, darkslategray66]
table {%
0 0.738601823708207
0 0.789260385005066
};
\addplot [line width=1.08pt, darkslategray66]
table {%
1 0.743498817966903
1 0.788416075650118
};
\addplot [line width=1.08pt, darkslategray66]
table {%
2 0.569739952718676
2 0.723404255319149
};
\addplot [line width=1.08pt, darkslategray66]
table {%
3 0.58983451536643
3 0.749408983451537
};
\end{axis}

\end{tikzpicture}}
%
\resizebox{\linewidth}{!}{
\begin{tikzpicture}

\definecolor{darkgray158150204}{RGB}{158,150,204}
\definecolor{darkslategray66}{RGB}{66,66,66}
\definecolor{dimgray85}{RGB}{85,85,85}
\definecolor{gainsboro229}{RGB}{229,229,229}
\definecolor{gray119}{RGB}{119,119,119}
\definecolor{indianred2049072}{RGB}{204,90,72}
\definecolor{steelblue69133171}{RGB}{69,133,171}
\definecolor{my_cp5_col6}{RGB}{2, 117, 216}
\definecolor{my_cp5_col5}{RGB}{92, 184, 92}
\definecolor{my_cp5_col4}{RGB}{91, 192, 222}
\definecolor{my_cp5_col3}{RGB}{240, 173, 78}
\definecolor{my_cp5_col2}{RGB}{217, 83, 79}
\definecolor{my_cp5_col1}{RGB}{52, 167, 255}
\begin{axis}[
axis background/.style={fill=gainsboro229},
axis line style={white},
tick align=outside,
tick pos=left,
x grid style={white},
xmin=-0.5, xmax=3.5,
xmajorticks=false,
xtick style={color=dimgray85},
xtick={0,1,2,3},
xticklabels={GrpA-1,GrpA-2,FCNN-1,FCNN-2},
y grid style={white},
 ylabel=\textcolor{dimgray85}{\LARGE \textcolor{white}{Uniform}},
ymajorgrids,
ymin=0.4, ymax=0.9,
ytick style={color=dimgray85}
]
\draw[draw=none,fill=my_cp5_col6,very thin] (axis cs:-0.4,0) rectangle (axis cs:0.4,0.637284701114488);
\draw[draw=none,fill=my_cp5_col5,very thin] (axis cs:0.6,0) rectangle (axis cs:1.4,0.678486997635934);
\draw[draw=none,fill=my_cp5_col4,very thin] (axis cs:1.6,0) rectangle (axis cs:2.4,0.587470449172577);
\draw[draw=none,fill=my_cp5_col3,very thin] (axis cs:2.6,0) rectangle (axis cs:3.4,0.567375886524823);
\addplot [line width=1.08pt, darkslategray66]
table {%
0 0.602836879432624
0 0.678824721377913
};
\addplot [line width=1.08pt, darkslategray66]
table {%
1 0.648936170212766
1 0.709219858156028
};
\addplot [line width=1.08pt, darkslategray66]
table {%
2 0.546099290780142
2 0.619385342789598
};
\addplot [line width=1.08pt, darkslategray66]
table {%
3 0.51418439716312
3 0.621749408983452
};
\end{axis}

\end{tikzpicture}}
\resizebox{\linewidth}{!}{
\begin{tikzpicture}

\definecolor{my_cp5_col6}{RGB}{2, 117, 216}
\definecolor{my_cp5_col5}{RGB}{92, 184, 92}
\definecolor{my_cp5_col4}{RGB}{91, 192, 222}
\definecolor{my_cp5_col3}{RGB}{240, 173, 78}
\definecolor{my_cp5_col2}{RGB}{217, 83, 79}
\definecolor{my_cp5_col1}{RGB}{52, 167, 255}
\definecolor{darkgray158150204}{RGB}{158,150,204}
\definecolor{darkslategray66}{RGB}{66,66,66}
\definecolor{dimgray85}{RGB}{85,85,85}
\definecolor{gainsboro229}{RGB}{229,229,229}
\definecolor{gray119}{RGB}{119,119,119}
\definecolor{indianred2049072}{RGB}{204,90,72}
\definecolor{steelblue69133171}{RGB}{69,133,171}

\begin{axis}[
axis background/.style={fill=gainsboro229},
axis line style={white},
tick align=outside,
tick pos=left,
x grid style={white},
xmin=-0.5, xmax=3.5,
xmajorticks=false,
xtick style={color=dimgray85},
xtick={0,1,2,3},
xticklabels={GrpA-1,GrpA-2,FCNN-1,FCNN-2},
y grid style={white},
ylabel=\textcolor{dimgray85}{\LARGE \textcolor{white}{Grid}},
ymajorgrids,
ymin=0.4, ymax=0.9,
ytick style={color=dimgray85}
]
\draw[draw=none,fill=my_cp5_col6,very thin] (axis cs:-0.4,0) rectangle (axis cs:0.4,0.724417426545086);
\draw[draw=none,fill=my_cp5_col5,very thin] (axis cs:0.6,0) rectangle (axis cs:1.4,0.788416075650118);
\draw[draw=none,fill=my_cp5_col4,very thin] (axis cs:1.6,0) rectangle (axis cs:2.4,0.544917257683215);
\draw[draw=none,fill=my_cp5_col3,very thin] (axis cs:2.6,0) rectangle (axis cs:3.4,0.595744680851064);
\addplot [line width=1.08pt, darkslategray66]
table {%
0 0.68693009118541
0 0.764969604863222
};
\addplot [line width=1.08pt, darkslategray66]
table {%
1 0.764775413711584
1 0.822695035460993
};
\addplot [line width=1.08pt, darkslategray66]
table {%
2 0.49645390070922
2 0.596926713947991
};
\addplot [line width=1.08pt, darkslategray66]
table {%
3 0.536643026004728
3 0.647754137115839
};
\end{axis}

\end{tikzpicture}}

	\end{subfigure}
	\caption{Binary Knot Classification Accuracy: In the left column, the number of samples is 125, where we can compare against \cite{finzi2020generalizing} as well as the FCNN. In the right column, the number of samples is 1000, which is too large for LieConv-1 and LieConv-2. Each row shows the distribution used to draw the samples.}
	\label{fig_knot_numsim}
\end{figure}
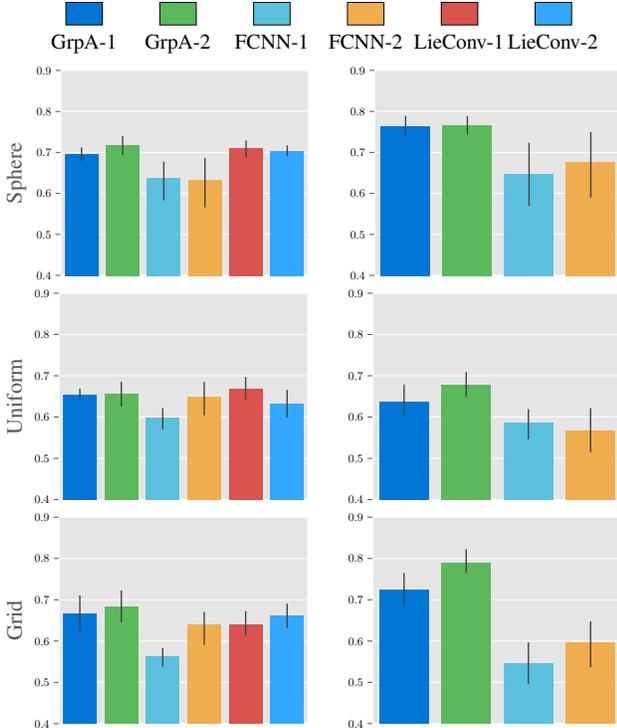

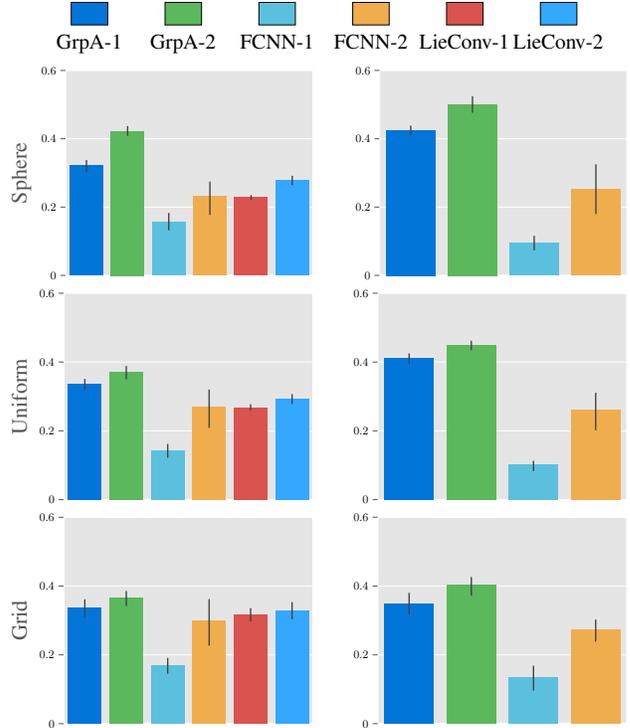
\begin{figure}[!ht]
	\begin{subfigure}{.5\textwidth}
	\centering




\definecolor{my_cp5_col6}{RGB}{2, 117, 216}
\definecolor{my_cp5_col5}{RGB}{92, 184, 92}
\definecolor{my_cp5_col4}{RGB}{91, 192, 222}
\definecolor{my_cp5_col3}{RGB}{240, 173, 78}
\definecolor{my_cp5_col2}{RGB}{217, 83, 79}
\definecolor{my_cp5_col1}{RGB}{52, 167, 255}

\usetikzlibrary{positioning,decorations.pathreplacing,shapes}


\def \scale {1}
\def \unit { \scale cm}

\def \vertsep {0.5*\scale}

\def \horzsep {1*\scale}


\tikzstyle{set} = [rectangle,color=black,
                    rounded corners = 0*\unit,
                    fill=black,
                    inner sep=0pt,
                    draw,
                    anchor = center,
                    line width=0.1mm]
                    
 
 
 \tikzstyle{myboxlabel} = [set,
fill=my_cp5_col3,
minimum width  = 0.485*\unit,
minimum height = 0.3*\unit]

\tikzstyle{dot} = [ circle,
                    minimum width  = 0.05*\unit,
                    fill=black,
                    color=black,
                    inner sep=0pt,
                    draw,
                    anchor = center ]


{\fontsize{8}{8}\selectfont

\begin{tikzpicture}[scale=\scale,rounded corners,ultra thick]


   
   \path (0,0) node [myboxlabel,fill=my_cp5_col6] (L1) {};
   \path (L1.south) ++ (0, 0) node [below, color=black] {GrpA-1};

   
    \path (L1.east)++(\horzsep,0) node [myboxlabel,fill=my_cp5_col5] (L2) {};
    \path (L2.south) ++ (0, 0) node [below, color=black] {GrpA-2};

     
     \path (L2.east)++(\horzsep,0) node [myboxlabel,fill=my_cp5_col4] (L3) {};
     \path (L3.south) ++ (0, 0) node [below, color=black] {FCNN-1};

         
     \path (L3.east)++(\horzsep,0) node [myboxlabel,fill=my_cp5_col3] (L4) {};
     \path (L4.south) ++ (0, 0) node [below, color=black] {FCNN-2};

      
      \path (L4.east)++(\horzsep,0) node [myboxlabel,fill=my_cp5_col2] (L5) {};
      \path (L5.south) ++ (0, 0) node [below, color=black] {LieConv-1};

      
      \path (L5.east)++(\horzsep,0) node [myboxlabel,fill=my_cp5_col1] (L6) {};
      \path (L6.south) ++ (0, 0) node [below, color=black] {LieConv-2};

%
%

\end{tikzpicture}

}
\end{subfigure}
\centering
	\centering
	\begin{subfigure}{.23\textwidth}
		\centering
		\resizebox{\linewidth}{!}{
\begin{tikzpicture}

\definecolor{my_cp5_col6}{RGB}{2, 117, 216}
\definecolor{my_cp5_col5}{RGB}{92, 184, 92}
\definecolor{my_cp5_col4}{RGB}{91, 192, 222}
\definecolor{my_cp5_col3}{RGB}{240, 173, 78}
\definecolor{my_cp5_col2}{RGB}{217, 83, 79}
\definecolor{my_cp5_col1}{RGB}{52, 167, 255}
\definecolor{burlywood231187113}{RGB}{231,187,113}
\definecolor{darkgray158150204}{RGB}{158,150,204}
\definecolor{darkslategray66}{RGB}{66,66,66}
\definecolor{dimgray85}{RGB}{85,85,85}
\definecolor{gainsboro229}{RGB}{229,229,229}
\definecolor{gray119}{RGB}{119,119,119}
\definecolor{indianred2049072}{RGB}{204,90,72}
\definecolor{steelblue69133171}{RGB}{69,133,171}
\definecolor{yellowgreen13817181}{RGB}{138,171,81}

\begin{axis}[
axis background/.style={fill=gainsboro229},
axis line style={white},
tick align=outside,
x grid style={white},
xmajorticks=false,
xmin=-0.5, xmax=5.5,
xtick style={color=dimgray85},
xtick={0,1,2,3,4,5},
xticklabels={GrpA-1,GrpA-2,FCNN-1,FCNN-2,LieConv-1,LieConv-2},
y grid style={white},
 ylabel=\textcolor{dimgray85}{\LARGE Sphere},
ymajorgrids,
ymin=0, ymax=0.6,
ytick pos=left,
ytick style={color=dimgray85}
]
\draw[draw=none,fill=my_cp5_col6,very thin] (axis cs:-0.4,0) rectangle (axis cs:0.4,0.320925110132159);
\draw[draw=none,fill=my_cp5_col5,very thin] (axis cs:0.6,0) rectangle (axis cs:1.4,0.422356828193833);
\draw[draw=none,fill=my_cp5_col4,very thin] (axis cs:1.6,0) rectangle (axis cs:2.4,0.155947136563877);
\draw[draw=none,fill=my_cp5_col3,very thin] (axis cs:2.6,0) rectangle (axis cs:3.4,0.231057268722467);
\draw[draw=none,fill=my_cp5_col2,very thin] (axis cs:3.6,0) rectangle (axis cs:4.4,0.227973568281938);
\draw[draw=none,fill=my_cp5_col1,very thin] (axis cs:4.6,0) rectangle (axis cs:5.4,0.278854625550661);
\addplot [line width=1.08pt, darkslategray66]
table {%
0 0.30352422907489
0 0.337555066079295
};
\addplot [line width=1.08pt, darkslategray66]
table {%
1 0.408480176211454
1 0.437007158590308
};
\addplot [line width=1.08pt, darkslategray66]
table {%
2 0.13182268722467
2 0.182714757709251
};
\addplot [line width=1.08pt, darkslategray66]
table {%
3 0.176960352422907
3 0.274672356828194
};
\addplot [line width=1.08pt, darkslategray66]
table {%
4 0.221137114537445
4 0.235134911894273
};
\addplot [line width=1.08pt, darkslategray66]
table {%
5 0.264424559471366
5 0.291646475770925
};
\end{axis}

\end{tikzpicture}}


  \resizebox{\linewidth}{!}{
\begin{tikzpicture}
\definecolor{my_cp5_col6}{RGB}{2, 117, 216}
\definecolor{my_cp5_col5}{RGB}{92, 184, 92}
\definecolor{my_cp5_col4}{RGB}{91, 192, 222}
\definecolor{my_cp5_col3}{RGB}{240, 173, 78}
\definecolor{my_cp5_col2}{RGB}{217, 83, 79}
\definecolor{my_cp5_col1}{RGB}{52, 167, 255}
\definecolor{burlywood231187113}{RGB}{231,187,113}
\definecolor{darkgray158150204}{RGB}{158,150,204}
\definecolor{darkslategray66}{RGB}{66,66,66}
\definecolor{dimgray85}{RGB}{85,85,85}
\definecolor{gainsboro229}{RGB}{229,229,229}
\definecolor{gray119}{RGB}{119,119,119}
\definecolor{indianred2049072}{RGB}{204,90,72}
\definecolor{steelblue69133171}{RGB}{69,133,171}
\definecolor{yellowgreen13817181}{RGB}{138,171,81}

\begin{axis}[
axis background/.style={fill=gainsboro229},
axis line style={white},
tick align=outside,
x grid style={white},
xmajorticks=false,
xmin=-0.5, xmax=5.5,
xtick style={color=dimgray85},
xtick={0,1,2,3,4,5},
xticklabels={GrpA-1,GrpA-2,FCNN-1,FCNN-2,LieConv-1,LieConv-2},
y grid style={white},
 ylabel=\textcolor{dimgray85}{\LARGE Uniform},
ymajorgrids,
ymin=0, ymax=0.6,
ytick pos=left,
ytick style={color=dimgray85}
]
\draw[draw=none,fill=my_cp5_col6,very thin] (axis cs:-0.4,0) rectangle (axis cs:0.4,0.335352422907489);
\draw[draw=none,fill=my_cp5_col5,very thin] (axis cs:0.6,0) rectangle (axis cs:1.4,0.3715859030837);
\draw[draw=none,fill=my_cp5_col4,very thin] (axis cs:1.6,0) rectangle (axis cs:2.4,0.141740088105727);
\draw[draw=none,fill=my_cp5_col3,very thin] (axis cs:2.6,0) rectangle (axis cs:3.4,0.269603524229075);
\draw[draw=none,fill=my_cp5_col2,very thin] (axis cs:3.6,0) rectangle (axis cs:4.4,0.26795154185022);
\draw[draw=none,fill=my_cp5_col1,very thin] (axis cs:4.6,0) rectangle (axis cs:5.4,0.293392070484581);
\addplot [line width=1.08pt, darkslategray66]
table {%
0 0.320154185022026
0 0.350881057268722
};
\addplot [line width=1.08pt, darkslategray66]
table {%
1 0.349666850220264
1 0.388215859030837
};
\addplot [line width=1.08pt, darkslategray66]
table {%
2 0.122020925110132
2 0.161569383259912
};
\addplot [line width=1.08pt, darkslategray66]
table {%
3 0.208689427312775
3 0.320154185022026
};
\addplot [line width=1.08pt, darkslategray66]
table {%
4 0.25880781938326
4 0.277422907488987
};
\addplot [line width=1.08pt, darkslategray66]
table {%
5 0.278191079295154
5 0.306720814977974
};
\end{axis}

\end{tikzpicture}}

\resizebox{\linewidth}{!}{
\begin{tikzpicture}
\definecolor{my_cp5_col6}{RGB}{2, 117, 216}
\definecolor{my_cp5_col5}{RGB}{92, 184, 92}
\definecolor{my_cp5_col4}{RGB}{91, 192, 222}
\definecolor{my_cp5_col3}{RGB}{240, 173, 78}
\definecolor{my_cp5_col2}{RGB}{217, 83, 79}
\definecolor{my_cp5_col1}{RGB}{52, 167, 255}
\definecolor{burlywood231187113}{RGB}{231,187,113}
\definecolor{darkgray158150204}{RGB}{158,150,204}
\definecolor{darkslategray66}{RGB}{66,66,66}
\definecolor{dimgray85}{RGB}{85,85,85}
\definecolor{gainsboro229}{RGB}{229,229,229}
\definecolor{gray119}{RGB}{119,119,119}
\definecolor{indianred2049072}{RGB}{204,90,72}
\definecolor{steelblue69133171}{RGB}{69,133,171}
\definecolor{yellowgreen13817181}{RGB}{138,171,81}

\begin{axis}[
axis background/.style={fill=gainsboro229},
axis line style={white},
tick align=outside,
x grid style={white},
xmajorticks=false,
xmin=-0.5, xmax=5.5,
xtick style={color=dimgray85},
xtick={0,1,2,3,4,5},
xticklabels={GrpA-1,GrpA-2,FCNN-1,FCNN-2,LieConv-1,LieConv-2},
y grid style={white},
 ylabel=\textcolor{dimgray85}{\LARGE Grid},
ymajorgrids,
ymin=0, ymax=0.6,
ytick pos=left,
ytick style={color=dimgray85}
]
\draw[draw=none,fill=my_cp5_col6,very thin] (axis cs:-0.4,0) rectangle (axis cs:0.4,0.337334801762115);
\draw[draw=none,fill=my_cp5_col5,very thin] (axis cs:0.6,0) rectangle (axis cs:1.4,0.364647577092511);
\draw[draw=none,fill=my_cp5_col4,very thin] (axis cs:1.6,0) rectangle (axis cs:2.4,0.16784140969163);
\draw[draw=none,fill=my_cp5_col3,very thin] (axis cs:2.6,0) rectangle (axis cs:3.4,0.299449339207048);
\draw[draw=none,fill=my_cp5_col2,very thin] (axis cs:3.6,0) rectangle (axis cs:4.4,0.316299559471366);
\draw[draw=none,fill=my_cp5_col1,very thin] (axis cs:4.6,0) rectangle (axis cs:5.4,0.329185022026432);
\addplot [line width=1.08pt, darkslategray66]
table {%
0 0.310132158590308
0 0.361247246696035
};
\addplot [line width=1.08pt, darkslategray66]
table {%
1 0.341949339207048
1 0.385691079295154
};
\addplot [line width=1.08pt, darkslategray66]
table {%
2 0.144810022026432
2 0.19108204845815
};
\addplot [line width=1.08pt, darkslategray66]
table {%
3 0.226756607929515
3 0.362026431718062
};
\addplot [line width=1.08pt, darkslategray66]
table {%
4 0.297026431718062
4 0.33557268722467
};
\addplot [line width=1.08pt, darkslategray66]
table {%
5 0.303741740088106
5 0.353309471365639
};
\end{axis}

\end{tikzpicture}}
		
	\end{subfigure}
	\begin{subfigure}{.23\textwidth}
		\centering
%
\resizebox{\linewidth}{!}{
\begin{tikzpicture}
\definecolor{my_cp5_col6}{RGB}{2, 117, 216}
\definecolor{my_cp5_col5}{RGB}{92, 184, 92}
\definecolor{my_cp5_col4}{RGB}{91, 192, 222}
\definecolor{my_cp5_col3}{RGB}{240, 173, 78}
\definecolor{my_cp5_col2}{RGB}{217, 83, 79}
\definecolor{my_cp5_col1}{RGB}{52, 167, 255}
\definecolor{darkgray158150204}{RGB}{158,150,204}
\definecolor{darkslategray66}{RGB}{66,66,66}
\definecolor{dimgray85}{RGB}{85,85,85}
\definecolor{gainsboro229}{RGB}{229,229,229}
\definecolor{gray119}{RGB}{119,119,119}
\definecolor{indianred2049072}{RGB}{204,90,72}
\definecolor{steelblue69133171}{RGB}{69,133,171}

\begin{axis}[
axis background/.style={fill=gainsboro229},
axis line style={white},
tick align=outside,
tick pos=left,
x grid style={white},
xmin=-0.5, xmax=3.5,
xmajorticks=false,
xtick style={color=dimgray85},
xtick={0,1,2,3},
xticklabels={GrpA-1,GrpA-2,FCNN-1,FCNN-2},
y grid style={white},
 ylabel=\textcolor{dimgray85}{ \LARGE \textcolor{white}{Sphere}},
ymajorgrids,
ymin=0, ymax=0.6,
ytick style={color=dimgray85}
]
\draw[draw=none,fill=my_cp5_col6,very thin] (axis cs:-0.4,0) rectangle (axis cs:0.4,0.425267463813719);
\draw[draw=none,fill=my_cp5_col5,very thin] (axis cs:0.6,0) rectangle (axis cs:1.4,0.500917767988253);
\draw[draw=none,fill=my_cp5_col4,very thin] (axis cs:1.6,0) rectangle (axis cs:2.4,0.0947136563876652);
\draw[draw=none,fill=my_cp5_col3,very thin] (axis cs:2.6,0) rectangle (axis cs:3.4,0.252019089574156);
\addplot [line width=1.08pt, darkslategray66]
table {%
0 0.410163624921334
0 0.438483322844556
};
\addplot [line width=1.08pt, darkslategray66]
table {%
1 0.476133443465492
1 0.524596182085169
};
\addplot [line width=1.08pt, darkslategray66]
table {%
2 0.0734214390602056
2 0.116005873715125
};
\addplot [line width=1.08pt, darkslategray66]
table {%
3 0.179515418502203
3 0.32507342143906
};
\end{axis}

\end{tikzpicture}}
%
\resizebox{\linewidth}{!}{
\begin{tikzpicture}
\definecolor{my_cp5_col6}{RGB}{2, 117, 216}
\definecolor{my_cp5_col5}{RGB}{92, 184, 92}
\definecolor{my_cp5_col4}{RGB}{91, 192, 222}
\definecolor{my_cp5_col3}{RGB}{240, 173, 78}
\definecolor{my_cp5_col2}{RGB}{217, 83, 79}
\definecolor{my_cp5_col1}{RGB}{52, 167, 255}
\definecolor{darkgray158150204}{RGB}{158,150,204}
\definecolor{darkslategray66}{RGB}{66,66,66}
\definecolor{dimgray85}{RGB}{85,85,85}
\definecolor{gainsboro229}{RGB}{229,229,229}
\definecolor{gray119}{RGB}{119,119,119}
\definecolor{indianred2049072}{RGB}{204,90,72}
\definecolor{steelblue69133171}{RGB}{69,133,171}

\begin{axis}[
axis background/.style={fill=gainsboro229},
axis line style={white},
tick align=outside,
tick pos=left,
x grid style={white},
xmajorticks=false,
xmin=-0.5, xmax=3.5,
xtick style={color=dimgray85},
xtick={0,1,2,3},
xticklabels={GrpA-1,GrpA-2,FCNN-1,FCNN-2},
y grid style={white},
 ylabel=\textcolor{dimgray85}{ \LARGE \textcolor{white}{Uniform}},
ymajorgrids,
ymin=0, ymax=0.6,
ytick style={color=dimgray85}
]
\draw[draw=none,fill=my_cp5_col6,very thin] (axis cs:-0.4,0) rectangle (axis cs:0.4,0.411264946507237);
\draw[draw=none,fill=my_cp5_col5,very thin] (axis cs:0.6,0) rectangle (axis cs:1.4,0.449155653450808);
\draw[draw=none,fill=my_cp5_col4,very thin] (axis cs:1.6,0) rectangle (axis cs:2.4,0.102422907488987);
\draw[draw=none,fill=my_cp5_col3,very thin] (axis cs:2.6,0) rectangle (axis cs:3.4,0.261380323054332);
\addplot [line width=1.08pt, darkslategray66]
table {%
0 0.395055852737571
0 0.424964600377596
};
\addplot [line width=1.08pt, darkslategray66]
table {%
1 0.434104258443466
1 0.461820851688693
};
\addplot [line width=1.08pt, darkslategray66]
table {%
2 0.0831451908957415
2 0.112885462555066
};
\addplot [line width=1.08pt, darkslategray66]
table {%
3 0.201541850220264
3 0.31057268722467
};
\end{axis}

\end{tikzpicture}}
\resizebox{\linewidth}{!}{
\begin{tikzpicture}
\definecolor{my_cp5_col6}{RGB}{2, 117, 216}
\definecolor{my_cp5_col5}{RGB}{92, 184, 92}
\definecolor{my_cp5_col4}{RGB}{91, 192, 222}
\definecolor{my_cp5_col3}{RGB}{240, 173, 78}
\definecolor{my_cp5_col2}{RGB}{217, 83, 79}
\definecolor{my_cp5_col1}{RGB}{52, 167, 255}
\definecolor{darkgray158150204}{RGB}{158,150,204}
\definecolor{darkslategray66}{RGB}{66,66,66}
\definecolor{dimgray85}{RGB}{85,85,85}
\definecolor{gainsboro229}{RGB}{229,229,229}
\definecolor{gray119}{RGB}{119,119,119}
\definecolor{indianred2049072}{RGB}{204,90,72}
\definecolor{steelblue69133171}{RGB}{69,133,171}

\begin{axis}[
axis background/.style={fill=gainsboro229},
axis line style={white},
tick align=outside,
tick pos=left,
x grid style={white},
xmin=-0.5, xmax=3.5,
xmajorticks=false,
xtick style={color=dimgray85},
xtick={0,1,2,3},
xticklabels={GrpA-1,GrpA-2,FCNN-1,FCNN-2},
y grid style={white},
 ylabel=\textcolor{dimgray85}{ \LARGE \textcolor{white}{Grid}},
ymajorgrids,
ymin=0, ymax=0.6,
ytick style={color=dimgray85}
]
\draw[draw=none,fill=my_cp5_col6,very thin] (axis cs:-0.4,0) rectangle (axis cs:0.4,0.3470736312146);
\draw[draw=none,fill=my_cp5_col5,very thin] (axis cs:0.6,0) rectangle (axis cs:1.4,0.402716593245228);
\draw[draw=none,fill=my_cp5_col4,very thin] (axis cs:1.6,0) rectangle (axis cs:2.4,0.133627019089574);
\draw[draw=none,fill=my_cp5_col3,very thin] (axis cs:2.6,0) rectangle (axis cs:3.4,0.273311306901615);
\addplot [line width=1.08pt, darkslategray66]
table {%
0 0.316236626809314
0 0.3802745437382
};
\addplot [line width=1.08pt, darkslategray66]
table {%
1 0.371696035242291
1 0.426225220264317
};
\addplot [line width=1.08pt, darkslategray66]
table {%
2 0.0962968061674009
2 0.168506791483113
};
\addplot [line width=1.08pt, darkslategray66]
table {%
3 0.238252569750367
3 0.302496328928047
};
\end{axis}

\end{tikzpicture}}

	\end{subfigure}
	\caption{ModelNet10 Classification Test Accuracy: In the left column, the number of samples is 125, and in the right column, the number of samples is 1000. Each row shows the distribution used to draw the samples. Similar to the knot dataset, we can only comapre against \cite{finzi2020generalizing} when the number of samples is small.}
	\label{fig_modelnet10_numsim}
\end{figure}

\section{Numerical Results}  \label{sec_numericals}
\vspace{-5pt}



We evaluate the Lie group algebra filters on two problems with $SO(3)$ group symmetries. In particular, we consider a custom binary classification for knots in $\mathbb{R}^3$ and the ModelPoint10 \cite{wu20153d} dataset.
%
%
%
%
%

{\bf{Knot Dataset:} }We introduce the problem of binary classification on knots. Specifically, we would like to distinguish between a Trefoil knot, characterized by $(x, y, z) = (\cos(t) + 2\cos(2t), \sin(t) - 2\sin(2t), -\sin(3t))$,
%
and the Listing's knot (also known as the figure-eight knot), characterized by $(x, y, z)$ $=((2+ \cos(2t))\cos(3t),$$ (2+ \cos(2t))$$\sin(3t),$$\sin(4t))$.
%
%
The knot is evaluated $N$ times evenly spaced between zero and $2\pi$. Signals are then rotated by selecting a group action $g\in SO(3)$ at random. We jitter the position of the points with Gaussian noise with zero mean and $\sigma$ standard deviation. 

An advantage to our approach is the ability to process data with an arbitrary sampling scheme. For this reason, we consider three distinct sets $X$ on which the knots are projected: \emph{Grid} has evenly spaced samples on the cube, \emph{Uniform} has samples drawn uniformly at random on the cube, and \emph{Sphere} has samples drawn uniformly at random in the sphere. 
%
To project the knot on each sampling scheme, we use the following procedure. For each point in the set $X$, we find the $k$ nearest neighbors from the knot. We compute the mean of these distances, and set the value of each sample $s\in X$ on the grid equal to
  $  \bbf'(s) = [\tau - \frac{1}{k}\sum_{x \in \textrm{KNN}(s)} \bbf(x)]_+,$
where $[\cdot]_+$ clips negative values and $\tau$ is some threshold. 

For our experiments, we set $N = 200$, and the Gaussian jitter position noise is drawn from $\ccalN(0,0.01)$ during training and $\ccalN(0,0.1)$ for testing, with $\tau=1$. There are 120 knots used during training and 140 knots used during testing. The balance of the classes are equal. 

{\bf ModelPoint10 Dataset \cite{wu20153d}:}
The ModelNet10 \cite{wu20153d} dataset contains 4,899 CAD models for 10 categories. The dataset is split into 3,991 samples for training and 908 for testing. After normalizing the dataset, we rotate the signals so they are no longer aligned. In particular, similar to the Knot dataset generation, signals are rotated at random by selecting a group action $g\in SO(3)$. We project each point cloud onto the Grid, Uniform, Sphere, and Gaussian sampling schemes described in the previous subsection.

{\bf Implementation details:}
We train each model for 100 iterations with a batchsize of 16. We train the models with cross entropy loss and an ADAM optimizer with no weight decay. For each dataset, we run each simulation ten times with different random seeds. For all simulations, we set the learning rate equal to $10^{-3}$. We evaluate the performance of our model with the classification accuracy on the test data and compare it to two baselines. The first is a naive fully connected neural network (FCNN) with one hidden layer of length 50. The second is a filter implementation of LieConv \cite{finzi2020generalizing}, where we remove the residual net architecture to compare the convolution layer itself. We denote our proposed filter by GrpA. For all methods, we consider one and two layer architectures, denoted by the number after the method in Figures \ref{fig_knot_numsim} and \ref{fig_modelnet10_numsim}.


{\bf Discussion:} The classification test accuracies on the Knot and ModelNet10 datasets are shown in Figures \ref{fig_knot_numsim} and \ref{fig_modelnet10_numsim} respectively. First, consider the trend of adding a second layer. With only a few exceptions, adding a second layer to the model increases the accuracy across all models, all sample sizes, and all sample schemes. This is expected due to the increased expressibility of the model with more parameters. Next, we look at the performance across the sample sizes. In general, for FCNN, increasing the sample size decreases the performance. In contrast, the proposed GrpA networks increase in accuracy when the sample size is increased. Note that the increased sample accuracy for LieConv is not computed due to intractability for larger dataset sizes. Finally, we also notice a difference in the test accuracy depending on the sampling method chosen. In particular, we find that the performance generally increases from Uniform to Grid to Sphere. The difference between Sphere and Grid is better highlighted in the ModelNet10 numerical simulations.

\vspace{-5pt}



\section{Conclusion and Future work} 
\vspace{-1pt}
In this work, we considered Lie group convolutions on non-homogeneous spaces by using the Lie group algebra homomorphism. In particular, we extended the ASP model to Banach $*$-algebras, proposed two algebraically justified relaxations, and established stability through a connection to multigraphs. Our numerical simulations revealed that the signal sampling plays a large role in the overall performance of \emph{all} models and that GrpA not only outperforms LieConv in most settings, but also enables group convolutions in large dimensional spaces through the use of sparse transformation matrices. We leave the formal investigation to bound the distance between the ideal and realized filter to future work, and we believe the bound can be achieved by considering the bandwidths of the signal and Lie group algebra.

\newpage 
\bibliographystyle{IEEEbib}
\bibliography{refs}

\end{document}